\documentclass[preprintnumbers,amsmath,amssymb,latexsym,array,enumerate,twocolumn,pra,showpacs]{revtex4-1}
\usepackage{graphicx}
\usepackage{dcolumn}
\usepackage{bm}
\usepackage{datetime}
\usepackage{hyperref}

\usepackage{amsmath}
\usepackage{amssymb}
\usepackage{enumitem}

\usepackage{braket}
\usepackage{mathrsfs}
\usepackage{amsthm}
\usepackage{slashed}
\usepackage{fancyhdr}
\usepackage{empheq}
\usepackage{mathtools}
\usepackage{qcircuit}
\usepackage{array}
\usepackage[caption=false]{subfig}

\newcommand{\del}{\boldsymbol{\nabla}}

\begin{document}

\preprint{INT-PUB-18-061}

\title{Oracles for Gauss's law on digital quantum computers}


\author{Jesse R.~Stryker}
\email[]{stryker@uw.edu}
\affiliation{Institute for Nuclear Theory, University of Washington, Seattle, Washington 98195-1550, USA}

\date{\today}

\begin{abstract}
  Formulating a lattice gauge theory using only physical degrees of freedom generically leads to nonlocal interactions.
  A local Hamiltonian is desirable for quantum simulation, and this is possible by treating the Hilbert space as a subspace of a much larger Hilbert space in which Gauss's law is not automatic.
  Digital quantum simulations of this local formulation will wander into unphysical sectors due to errors from Trotterization or from quantum noise.
  In this work, oracles are constructed that use local Gauss law constraints to projectively distinguish physical and unphysical wave functions in Abelian lattice gauge theories.
  Such oracles can be used to detect errors that break Gauss's law.
\end{abstract}


\maketitle

\section{\label{sec:intro}Introduction}

Quantum devices are expected to efficiently simulate quantum mechanical many-body problems such as field theories
\cite{
  benioffComputerPhysical80,
  feynmanSimulatingPhysics82,
  lloydUniversalQuantum96,
  jordan.lee.eaQuantumAlgorithms12%
},
in particular lattice gauge theories (LGTs)
\cite{
  byrnes.yamamotoSimulatingLattice06,
  delascuevas.dur.eaUnifyingAll09,
  stannigel.hauke.eaConstrainedDynamics14,
  mezzacapo.rico.eaNonAbelianSU15,
  martinez.muschik.eaRealtimeDynamics16,
  zohar.cirac.eaQuantumSimulations16,
  dalmonte.montangeroLatticeGauge16,
  zohar.farace.eaDigitalLattice17,
  zohar.farace.eaDigitalQuantum17,
  zohar.reznikConfinementLattice11,
  banerjee.dalmonte.eaAtomicQuantum12,
  zohar.cirac.eaQuantumSimulations13,
  bazavov.meurice.eaGaugeinvariantImplementation15,
  kasper.hebenstreit.eaImplementingQuantum17,
  zhang.unmuth-yockey.eaQuantumSimulation18,
  bender.zohar.eaDigitalQuantum18,
  zache.hebenstreit.eaQuantumSimulation18,
  klco.dumitrescu.eaQuantumclassicalComputation18,
  hackett.howe.eaDigitizingGauge18,
  raychowdhury.strykerTailoringNonAbelian18%
}.
Except for small systems, such problems are intractable with classical computing due to the exponential growth of resources required.
Some basic properties of field theories can be calculated efficiently using classical computation---in lattice quantum chromodynamics (QCD) these include hadron masses and resonances
\cite{briceno.dudek.eaScatteringProcesses18},
magnetic moments
\cite{parreno.savage.eaOctetBaryon17},
parton distribution functions
\cite{lin.nocera.eaPartonDistributions18}, 
and spectra and properties of light nuclei
\cite{
  nplqcdcollaboration.beane.eaLightNuclei13,
  nplqcdcollaboration.chang.eaMagneticStructure15,
  nplqcdcollaboration.savage.eaProtonProtonFusion17,
  nplqcdcollaboration.tiburzi.eaDoubleEnsuremath17%
}
---but generically the cost of calculations grows prohibitively fast. 
Nonzero density and real-time phenomena formulated in terms of path integrals, for example, are plagued by exponentially hard sign problems
\cite{
  fodor.katzNewMethod02,cohenFunctionalIntegrals03,
  kratochvila.deforcrandQCDSmall05,
  troyer.wieseComputationalComplexity05,
  schmidtLatticeQCD06,
  splittorff.verbaarschotPhaseFermion07,
  aartsCanStochastic09,
  aurorasciencecollaboration.cristoforetti.eaNewApproach12,
  cohen.gull.eaTamingDynamical15,
  alexandru.basar.eaSchwingerKeldyshFormalism17,
  alexandru.bedaque.eaFinitedensityMonte18%
}.
Thus, computing many important nonperturbative processes in QCD, such as hadronization in high-energy collisions, is intractable.

\emph{Ab initio} dynamical studies of LGTs would be made possible by mapping the dynamics onto the evolution of the state of a quantum device.
Hamiltonian LGT is conventionally formulated in terms of gauge and electric field operators living on the links of a spatial lattice
\cite{
  kogut.susskindHamiltonianFormulation75,
  creutzQuarksGluons84,
  smitIntroductionQuantum02%
},
with matter degrees of freedom living on the sites.
Time discretization has been removed by gauge-fixing to $A_0 = 0$ and taking the continuous time limit.
Electric fields generate gauge transformations, and it is convenient to use a basis in which they are diagonal.
The residual freedom of spatial gauge transformations means there are redundant or unphysical degrees of freedom;
in this formulation they show up as electric field configurations inconsistent with Gauss's law
\cite{
  lagrangeAttractionSpheroides73,
  gaussTheoriaAttractionis13,
  kogut.susskindHamiltonianFormulation75,
  creutzQuarksGluons84,
  smitIntroductionQuantum02%
}.
Violating Gauss's law means spoiling gauge invariance.

Implementing Gauss's law poses a technical challenge for both analog and digital quantum devices.
Analog quantum simulators are quantum systems whose low-energy degrees of freedom are engineered to mimic the low-energy regime of the target theory.
Digital quantum simulation involves mapping the Hilbert space of the target theory onto the discrete Hilbert space of a quantum computer (often an array of qubits) and effecting the time evolution operator by a sequence of gates.
This paper is concerned with simulation on digital quantum computers, which can realize universal quantum computation and fault tolerance.

Ideally the states of a quantum computer would represent only distinct, physical states, since qubits are otherwise wasted simulating the unphysical ones, but such formulations generically lead to nonlocal and less symmetric interactions that are difficult to simulate efficiently.
However, by keeping the states that violate Gauss's law, only an exponentially small fraction of the qubit configurations correspond to physical states
\footnote{
  The ``unphysical'' sectors may alternatively be viewed as different static (external) charge configurations, which are separated by a superselection rule.
  In this paper, the ``physical'' subspace refers to that sector having no static charges anywhere, but the results could be straightforwardly adapted to other superselection sectors.
}.
While time evolution of an initial state that is physical ought to preclude the need to worry about unphysical sectors, quantum noise and systematic errors introduced by approximations can generate components along the many unphysical directions.
Previous work has addressed these constraints
by using engineered classical noise \cite{stannigel.hauke.eaConstrainedDynamics14},
by living with the unphysical sectors on the premise their effect is reduced with decreasing time-step size
\cite{
  mezzacapo.rico.eaNonAbelianSU15%
},
and by mapping purely physical degrees of freedom onto the quantum computer
\cite{
  martinez.muschik.eaRealtimeDynamics16,
  klco.dumitrescu.eaQuantumclassicalComputation18
}.
More approaches to Gauss's law have also been investigated for analog simulators
\cite{
  zohar.reznikConfinementLattice11, 
  banerjee.dalmonte.eaAtomicQuantum12,
  zohar.cirac.eaQuantumSimulations13,
  bazavov.meurice.eaGaugeinvariantImplementation15,
  kasper.hebenstreit.eaImplementingQuantum17,
  zhang.unmuth-yockey.eaQuantumSimulation18%
}.

The approach introduced herein develops algorithms to projectively measure physicality of states in Abelian LGTs by using constraint-checking oracles:
a state $\ket{\Psi}= \cos(\theta)\ket{\Psi_{\text{phys}}} + \sin(\theta)\ket{\Psi_{\text{unphys}}}$ will collapse to either $\ket{\Psi_{\text{phys}}}$ or $\ket{\Psi_{\text{unphys}}}$.
This could be useful as a method of filtering out unphysical errors---say, as an accept-reject step in a simulation (see also 
\cite{
    mcardle.yuan.eaErrorMitigated18,
    bonet-monroig.sagastizabal.eaLowcostError18%
}),
or as a subroutine for curing states afflicted with unphysical errors---without the need for all the qubits of fault-tolerant computation.
Rejecting unphysical states would be useful when working with small lattices, although large lattices presumably require a more sophisticated solution.
Non-Abelian generalizations are also underway
\cite{raychowdhury.strykerTailoringNonAbelian18}.

In \S\ref{sec:bg}, the needed gauge theory background is summarized.
Then, in \S\ref{sec:alg}, a procedure is given for constructing oracles that projectively distinguish physical states from unphysical ones.
In \S\ref{sec:ex}, constructions are shown in example theories.
Additional remarks on implementation follow since the examples omit details about the exact gates that would be programmed to a device.
Section \ref{sec:disc} then expands on the significance of constraint-checking circuits and their potential applications.

\section{\label{sec:bg}Mapping of $\mathbb{Z}_N$ and U(1)}
This section summarizes the Abelian LGT structure to be mapped onto a quantum computer.
The gauge groups considered are $g=\mathbb{Z}_N$ and $g=\text{U(1)}$.
Space is discretized on a cubic lattice $L$ with sites $s$.
For simplicity, $L$ is given periodic boundary conditions.
The lattice links $\ell\in L$ are associated with independent gauge field Hilbert spaces $\mathcal{H}_{\ell}$, each having an identical discrete basis:
\begin{equation}
  \braket{ m^\prime | m} = \delta_{m^\prime m} \ , \qquad \hat{1} = \sum_m \ket{m} \bra{m} \ ,
  \label{<+label+>}
\end{equation}
where
\begin{equation}
  m^\prime, m \in \begin{cases}\mathbb{Z}_N \ ,& \quad\text{if $g=\mathbb{Z}_N$} \ , \\
  \mathbb{Z} \ , & \quad\text{if $g=\text{U(1)}$}\ . \end{cases}
  \label{<+label+>}
\end{equation}
(The same symbol $\mathbb{Z}_N$ will be used for the integers modulo $N$ and for the $N^{\text{th}}$ roots of unity.)
The Hamiltonian is a function of link operators $\hat{U}_{\ell}$ associated with the links $\ell$ and their (dimensionless) conjugate electric fields $\hat{E}_{\ell}$, 
\begin{equation}
  \hat{U}_{\ell} = \sum_{m_{\ell}} \ket{m_{\ell}+1}\bra{m_{\ell}}, \quad \hat{E}_{\ell} =  \sum_{m_{\ell}} \ket{m_{\ell}} m_{\ell} \bra{m_{\ell}} \ .
  \label{<+label+>}
\end{equation}
For $g=\mathbb{Z}_N$, the electric field is periodic and the Hamiltonian really depends on its exponentiated form $\hat{Q}_{\ell}$,
\begin{align}
  \label{eq:zNperiodicity}
  \ket{m_{\ell}} &\equiv \ket{m_{\ell} \ \text{(mod $N$)} }  \ ,\\ 
  \label{eq:zNcharge}
  \quad \hat{Q}_{\ell} &\equiv e^{ 2\pi i\hat{E}_{\ell} /N} \\
  \notag
  &=  \sum_{m_{\ell}=0}^{N-1} \ket{m_{\ell}} e^{2\pi i m_{\ell} /N} \bra{m_{\ell}}  \ . 
\end{align}
Operators associated with different links commute, and the same-link commutation relations are
\begin{align}
  \hat{Q}_{\ell} \hat{U}_{\ell} \hat{Q}_{\ell}^\dagger &=  \hat{U}_{\ell} e^{2\pi i /N} \ ,\quad \text{ if $g=\mathbb{Z}_N$} \ ,\\
  [\hat{E}_{\ell} , \hat{U}_{\ell} ] &= \hat{U}_{\ell} \ ,\qquad\qquad \text{if $g=\text{U(1)}$} \ .
  \label{eq:abelianCR}
\end{align}
Link labels $\ell$ will now only be displayed when necessary.

For adding in matter there are many possibilities;
to keep things simple yet illustrative, I consider matter fields with anticommuting statistics and carrying unit charge.
Each matter species is labeled with a collective index $\sigma$, which could include flavor or Dirac indices, and has possible occupation numbers $0\leq n_\sigma \leq 1$.
The results of this paper are easily extended to other possibilities.

A lattice ``configuration'' or ``basis state'' will be used to refer to any state $\ket{\mathbf{E},\rho}$ with definite electric fields on the links and occupation numbers on the sites.
Usually only one site $s\in L$ is under consideration, so the quantum numbers associated with other sites or links not attached to $s$ are suppressed,
\begin{equation*}
  \ket{ \mathbf{E} , \rho }  \rightarrow \otimes_{i=1}^D \ket{E_{i}(s)} \otimes_{i=1}^D \ket{E_{i}(s-\hat{e}_i)} \otimes_\text{$\sigma$} \ket{n_{\sigma}(s)} 
  \label{<+label+>}
\end{equation*}
\footnote{Technically, an ordering must be specified to make the tensor product of states in a fermionic Fock space well-defined.
  It is assumed that this has been done.
Fermionic statistics are otherwise irrelevant to the oracles.}.
``Physicality'' of a state will refer to the Gauss law constraint being satisfied at each site $s$.
It is convenient to express this by using Gauss law and physicality operators $\hat{G}_s$ and $\hat{F}_s$, 
\begin{align}
  \begin{split}
    \hat{G}_s  &\equiv(\del \cdot \hat{\mathbf{E}})(s)  - \hat{\rho}(s) \label{eq:G_s}\\
    &= \sum_{i=1}^D (\hat{E}_{i}(s) - \hat{E}_{i}(s-\hat{e}_i)  ) - \sum_{\sigma} e_{\sigma} \hat{n}_{\sigma}(s) \ ,
  \end{split} \\
  \hat{F}_s &\equiv \begin{cases}  \sum\limits_{k=0}^{N-1} \frac{1}{N}e^{-2\pi i k \hat{G}_s / N } \ , & \text{if $g=\mathbb{Z}_N$} , \\[2ex] \int_0^{2\pi} \frac{d\phi}{2\pi} e^{-i \phi\hat{G}_s} \ , & \text{if $g=\text{U(1)}$}  .\end{cases}
    \label{eq:F_s}
\end{align}
Above, $\del \cdot $ is a discrete divergence operator and  $e_{\sigma}=\pm1$ are charges.
$\hat{F}_s$ simply projects onto the subspace of configurations with Gauss's law satisfied at $s$,
\begin{eqnarray}
  \hat{F}_s \ket{\text{phys}} &=& \ket{\text{phys}} \ ,\\
  \hat{F}_s \ket{\text{unphys}} &=& 0 \ .
  \label{<+label+>}
\end{eqnarray}
The notation $F_s(\mathbf{E},\rho)$ will be used for the eigenvalue of $\ket{\mathbf{E},\rho}$ with respect to $\hat{F}_s$.
The operators $\hat{F}_s$ are spatially local, but a full lattice configuration is only truly physical if $F_s(\mathbf{E},\rho)=1$ for all $s\in L$.
Oracle operators $\hat{O}_s$ can now be introduced in terms of the physicality operators,
\begin{equation}
  \hat{O}_s = \hat{1}- 2\hat{F}_s = e^{i \pi \hat{F}_s }\ .
  \label{<+label+>}
\end{equation}
The oracle ``flags'' states that satisfy Gauss's law at $s$,
\begin{equation}
    \hat{O}_s  \ket{ \mathbf{E} , \rho } =  \ket{ \mathbf{E} , \rho } (-)^{F_s (\mathbf{E}, \rho )}\ . 
  \label{<+label+>}
\end{equation}
Operators with such behavior are closely related to Grover's quantum search algorithm 
\cite{
  groverQuantumMechanics97,
  nielsen.chuangQuantumComputation11%
}.
Understanding $\hat{O}_s$ as a quantum circuit that computes or checks constraints is a central task in the remaining sections.

Digital quantum simulation requires a Hilbert space of finite dimensionality due to the finite number of qubits (or more generally, qudits).
For $g=\mathbb{Z}_N$, a finite lattice volume automatically renders the gauge Hilbert space finite-dimensional.
For $g=\text{U(1)}$, the links' electric fields are uniformly truncated as well. 
Anticipating the use of qubits, the local link dimension for either $g$  is fixed to
$\dim \mathcal{H}_{{\ell}} = 2^{n}$,
and the states are labeled by non-negative integers $0 \leq \epsilon \leq 2^{n} -1$.
These correspond to some range of uniformly-spaced electric field values $E$, the linear relationship being
\begin{equation}
  \epsilon = E - E_{\text{min}}\ , \qquad E_{\text{max}} = E_{\text{min}} + 2^{n}-1 \ .
  \label{eq:eFieldMapping}
\end{equation}
The truncation when $g=\text{U(1)}$ renders the link operators $\hat{U},\hat{U}^\dagger $ nonunitary since they can destroy states at the ends of the ladder.
Crucially, the commutation relation $[\hat{E},\hat{U}]=\hat{U}$ survives truncation, and all of the formalism introduced above carries over straightforwardly.

\section{\label{sec:alg} Oracles for constraints}

In this section the oracle $\hat{O}_s$ is given as a quantum algorithm that internally checks the constraint (\ref{eq:G_s}) for computational basis states and applies a conditional phase flip.
A basic familiarity with quantum computation and the quantum circuit model is assumed;
however, nonpractitioners may refer to Appendix \ref{app:notation} for an introduction to most of the notation used in this paper.
To state an algorithm for constructing $\hat{O}_s$, the formalism of the previous section will be tailored to a quantum computer.

A binary representation for electric fields $\ket{\epsilon}$ is used.
The number of qubits per link is $n$, so the local link Hilbert space takes the form
$\mathcal{H}_{\ell} = \{\ket{0},\ket{1}\}^{\otimes n}$,
with $\dim\mathcal{H_{\ell}} = 2^n$.
The bit strings represented by the computational basis states are regarded as binary expressions for the electric field labels $0\leq\epsilon\leq (2^n-1)$.
$\epsilon_i(s)$ and $\epsilon_i(s-\hat{e}_i)$ are referred to as $\epsilon_i^{\text{OUT}}$ and $\epsilon_i^{\text{IN}}$.
And for simplicity, the occupation numbers $n_\sigma=0,1$ will be identified with the computational labels 0 and 1.

The constraint function (\ref{eq:G_s}) at $s$ can be rewritten as
\begin{align}
  \label{eq:chargedConstraint}
  \begin{split}
    \del \cdot \mathbf{E} - \rho  =& \left( \sum_{i=1}^D \epsilon_{i}^{\text{ OUT}} + \sum_{\sigma :\ e_{\sigma} <0} \ n_{\sigma} \right)\\
    &- \left( \sum_{i=1}^D \epsilon_{i}^{\text{ IN}} + \sum_{\sigma :\ e_{\sigma} >0}  \ n_{\sigma} \right) \ .
  \end{split}
\end{align}
This form suggests the negative and positive charges be separately absorbed into the internal computation of the out-flux and in-flux.
One can easily imagine a variety of ways to arrange the internal arithmetic to compute the constraint, but with the procedure outlined below there are cheap ways to include small numbers of fermions if the addition subroutines use incoming ``carry'' qubits.

The main ideas of the algorithm can be illustrated by constructing the oracle and entangling it with an auxiliary ``query bit'' $\ket{q}$  that gets flipped if Gauss's law holds at site $s$ for a given wave function.
(``Bit'' will frequently be used in place of ``qubit.'')
$\ket{q}$ acts as an observable probe for the oracle's action.
The procedure is as follows:
\begin{enumerate}
  \item Initialize needed work bits.
    This includes the query bit $\ket{q}$ in a $\hat{Z}$ basis state $\ket{0}$ or $\ket{1}$.
  \item Apply a Hadamard gate to the query bit.
  \item Compute both sums on the right-hand side of (\ref{eq:chargedConstraint}).
  \item Compute the difference of (or compare) these sums.
  \item Apply a phase flip if and only if the constraint vanishes and the query bit is set to $\ket{1}$.
  \item Undo (uncompute) the arithmetic of steps 4 and 3, restoring the state's original quantum numbers.
  \item Apply another Hadamard gate to the query bit.
  \item Measure the query bit in the $\hat{Z}$ basis.
\end{enumerate}
A flip of the query bit $\ket{q}\rightarrow \ket{q\oplus 1}$ is found if and only if Gauss's law is satisfied at the site. 
In practice, it is usually cheaper to forego evaluating $G_s$ proper (step 4) in favor of a more direct comparison of two integers.
\begin{figure}
  \includegraphics[width=.4\textwidth]{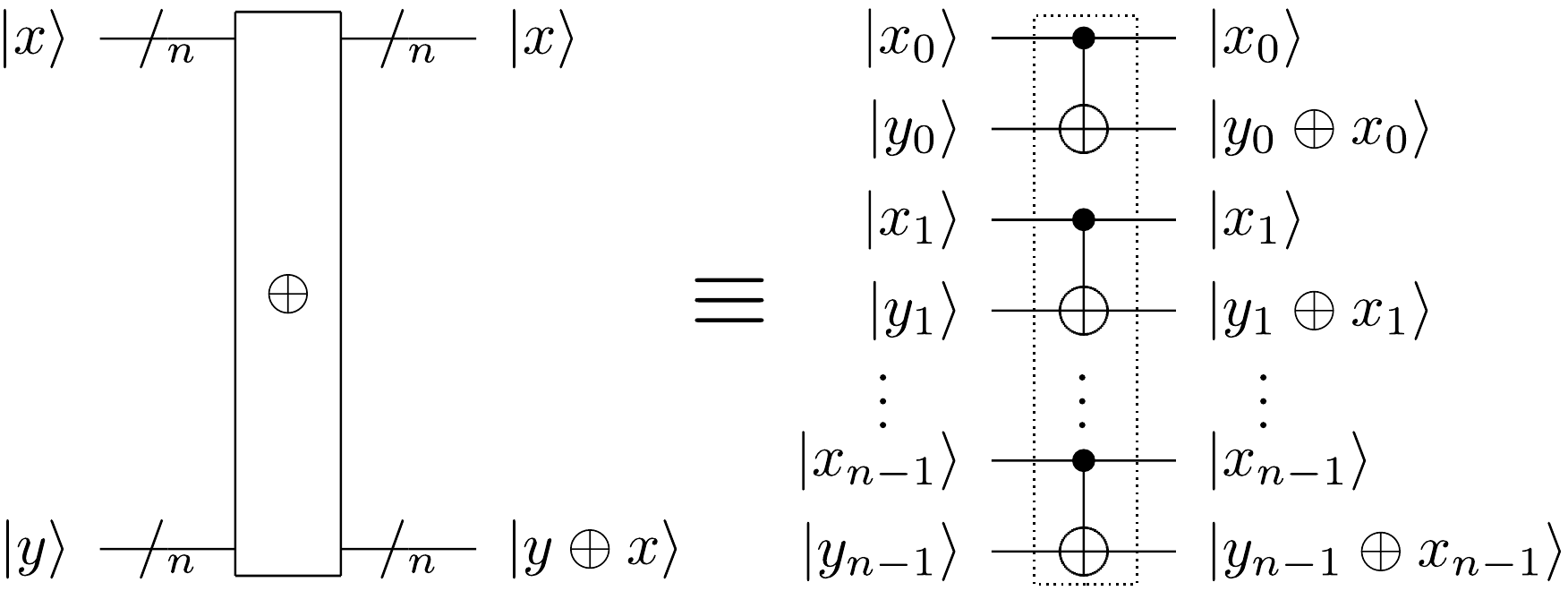}
   \caption{
     A multi-qubit subroutine $\oplus$ for comparing two $n$-bit integers $x$ and $y$ by adding all their corresponding bits modulo two.
     A (non)zero result for $y\oplus x$  means $x$ is (not) equal to $y$.
     \label{fig:CNOTs}}
\end{figure}
\begin{figure}
  \includegraphics[height=2cm]{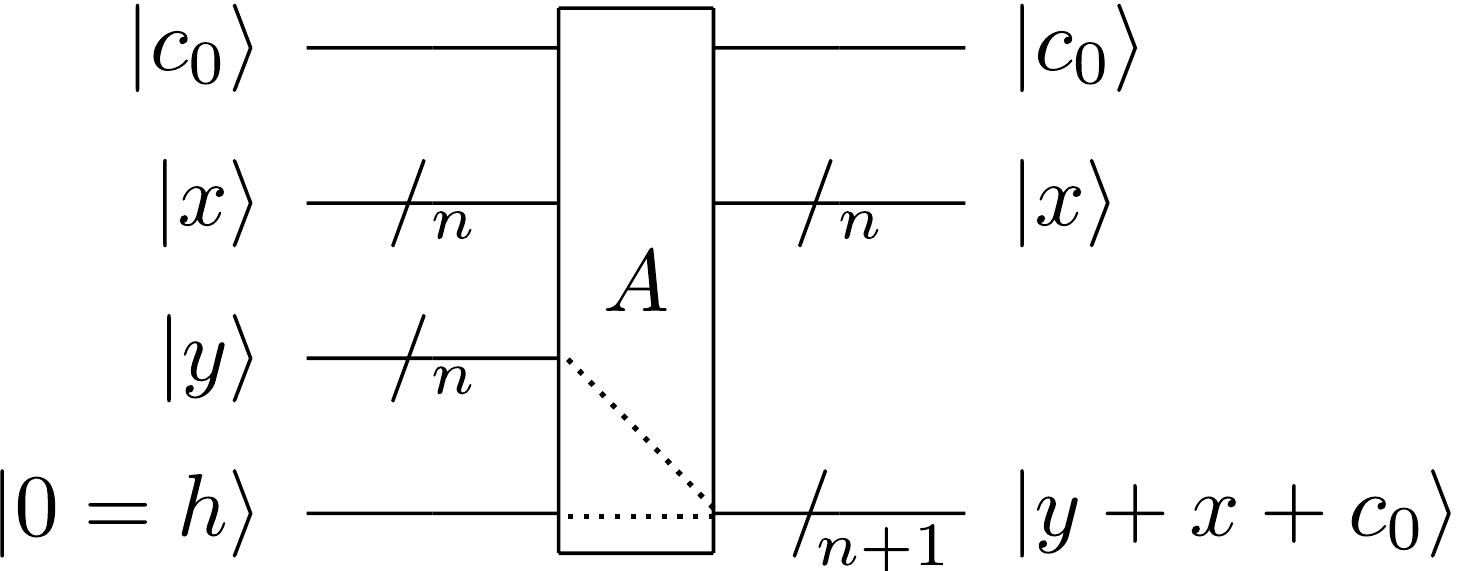}
   \caption{
     A generic adder $A$ for adding two $n$-bit integers ``in place.''
     $A$ is assumed to take an incoming carry bit $c_0$(=0,1), whose value can be added to $y+x$ at no additional cost.
     The overflow bit $h=0$ is needed to express the $(n+1)$-bit sum.
     \label{fig:add}}
\end{figure}
This is accomplished with a string of controlled-\textsc{not} gates, \textsc{cnot}s, as in Fig.~\ref{fig:CNOTs}.
These \textsc{cnot}s are denoted by a multi-qubit gate, $\oplus$.

Calls to the oracle will generally be described in terms of their action on computational basis states.
When the input lattice wave function is any superposition of physical ($F_s=1$) and unphysical ($F_s=0$) components, measuring the query bit afterward will probabilistically project the state onto one eigenspace of $\hat{F}_s$ or the other.

\section{\label{sec:ex}Examples and remarks}
\begin{figure*}
  \subfloat[  \label{fig:1DCharged}]{%
       \includegraphics[width=.72\textwidth]{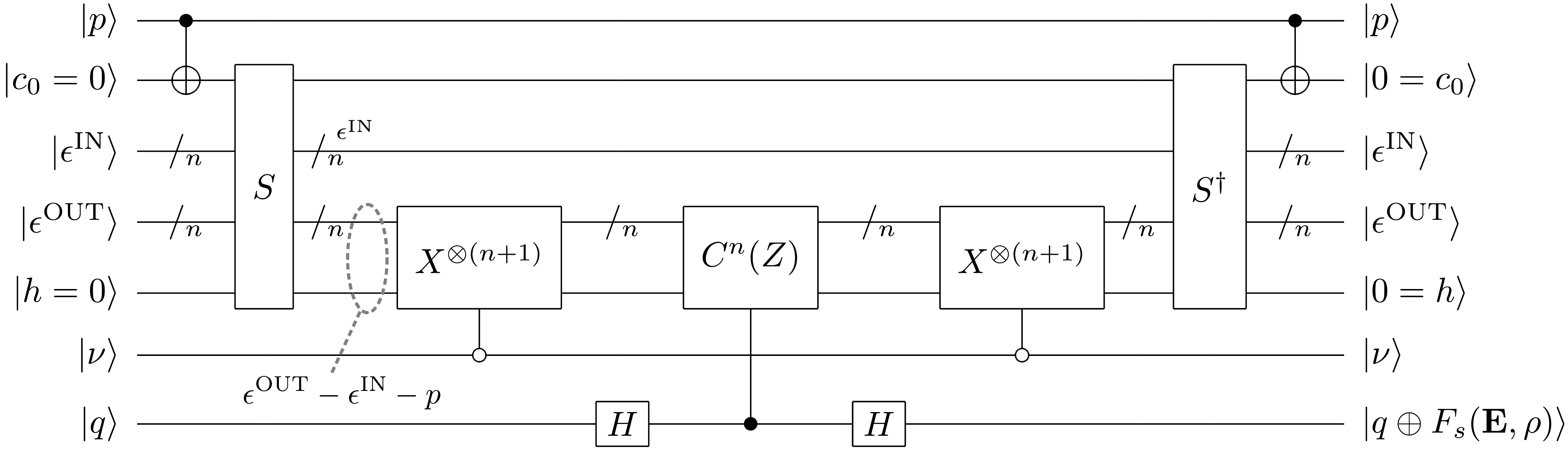}%
}\hfill\subfloat[ \label{fig:subtractor}]{%
  \includegraphics[width=.23\textwidth]{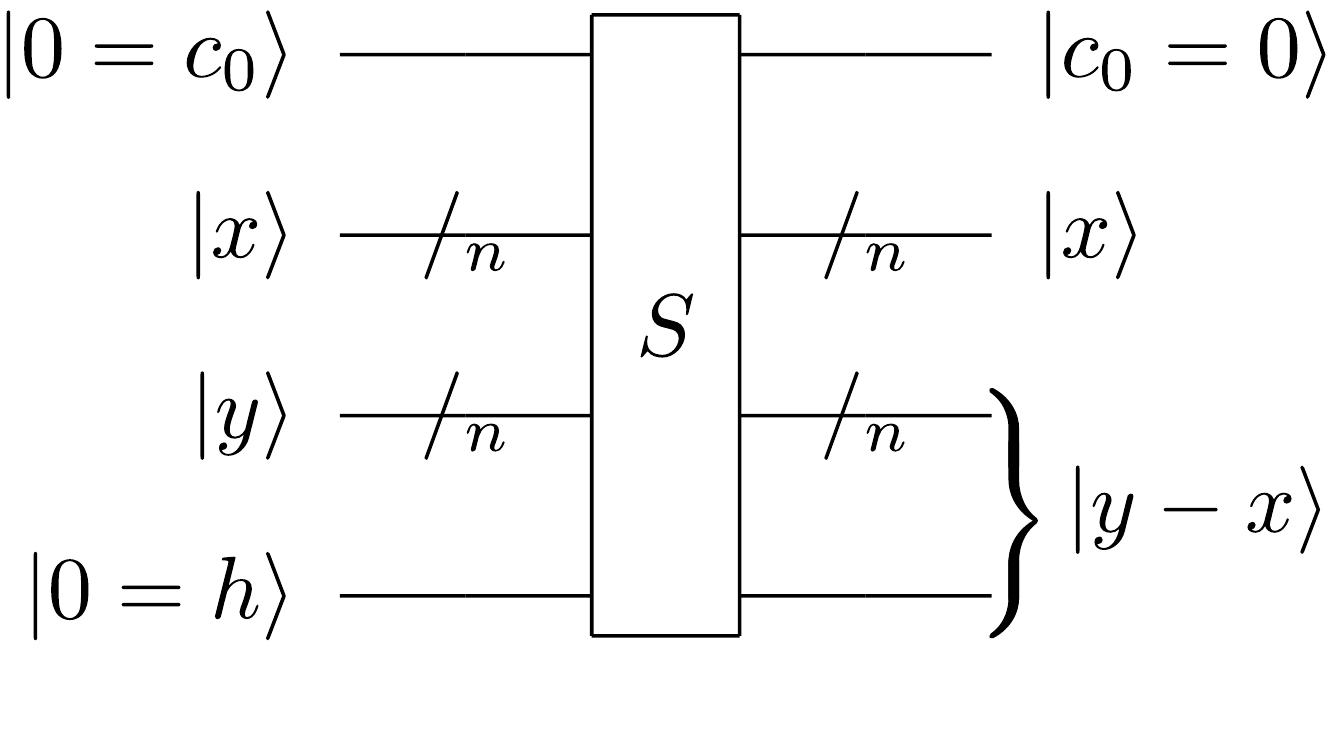}%
}
\caption{\label{fig:gaussOracle1DCharged}
  A query to the oracle for a U(1) or $\mathbb{Z}_{2^n}$ gauge theory in 1D.
  (a) The circuit for checking Gauss's law, which accommodates one Dirac fermion.
  (b) A subtractor routine $S$, which may be thought of as a modified version of $A$.
}
\end{figure*}
\begin{figure*}
  \begin{center}
    \includegraphics[width=.8\textwidth]{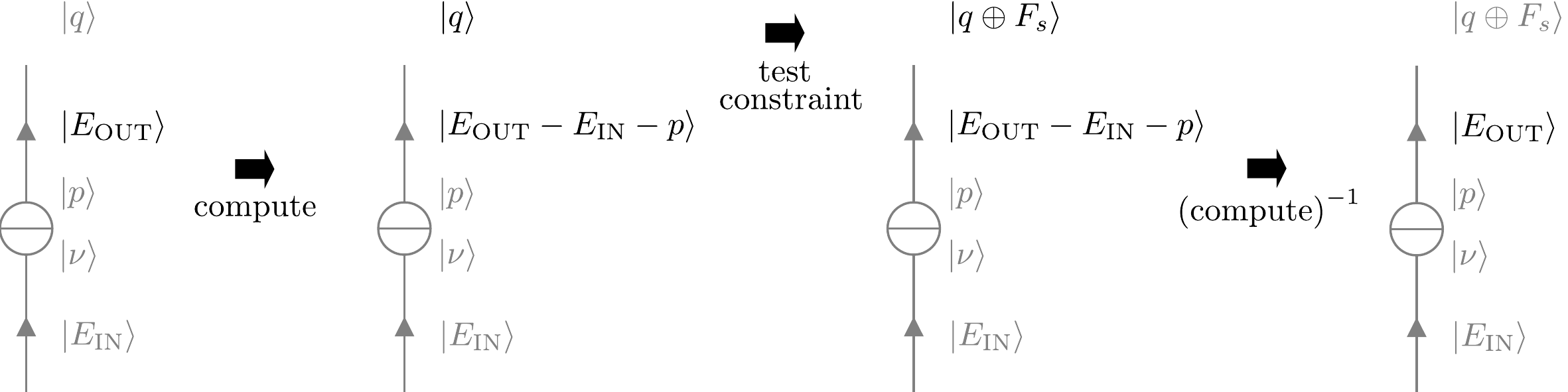}
    \caption{
      A schematic summary of how the 1D Gauss law circuit (Fig. \ref{fig:gaussOracle1DCharged}) acts on a basis state.
  }
  \label{fig:1DTrans}
\end{center}
\end{figure*}
This section expands on how the oracles work by showing circuits for example theories.
Truncated U(1) shares more features with the simulation of non-Abelian Lie groups than does $\mathbb{Z}_{2^n}$, so figures are shown for the former case and the small modifications needed for the latter are explained in the text.
The adders needed for checking a constraint are indicated by multi-qubit gates $A$, defined in Fig.~\ref{fig:add}.
It is common for addition algorithms to take an incoming carry bit $c_0$ initialized to zero, which can be exploited to account for fermions. 
For now, the adders' inner workings will be neglected because the exact choice of algorithm they use is mostly irrelevant to the oracle.
The simplest interesting matter scenarios will be seen to cost essentially the same as their pure gauge counterparts [except in one spatial dimension (1D)], but more general matter content requires introducing more adder subroutines---thereby increasing the number of required gates.
Thereafter, remarks are made about the adders and T gate counts are given.
\subsection{1D U(1) and $\mathbb{Z}_N$}
Figure \ref{fig:gaussOracle1DCharged} depicts an oracle for 1D U(1) gauge theory coupled to one Dirac fermion, while a schematic description of the circuit's actions is given in Fig. \ref{fig:1DTrans}.
The physical inputs from a lattice site are two $n$-bit electric field states $\ket{\epsilon^{\text{OUT,IN}}}$, and occupation numbers $\ket{\nu}$ and $\ket{p}$ corresponding to the negative and positive charges, respectively.
The incoming carry bit $\ket{c_0}$ is set equal to $\ket{p}$.
(One could simply use $\ket{p}$ as the carry bit.)
An overflow bit $\ket{h}$ is used for storing the result of the subtraction routine $S$ [Fig. \ref{fig:subtractor}].
This is the only example in which actually performing the subtraction in (\ref{eq:chargedConstraint}) does not cost more gates than a simple comparison.
Additional work bits could be needed by the subtraction subroutine $S$, or by the controlled $Z$.

In more detail:
The subtractor $S$, regarded as a modified version of $A$, first computes the difference of two $n$-bit electric fields, $\epsilon^{\text{OUT}}-\epsilon^{\text{IN}}$, using the relation
\begin{align}
  \label{eq:binarySub}
  a - b &= \overline{\bar{a} + b } \ , \\
  \label{eq:complement}
  \ket{\bar{a}}&=X^{\otimes n}\ket{a} \ .
\end{align}
The bar notation defined in (\ref{eq:complement}) indicates the ``ones' complement'' of $a$.
However, this notation is abused in (\ref{eq:binarySub});
the precise meaning is
\begin{align}
  \ket{\underbrace{a}_{n\text{ bits}}-\underbrace{b}_{n\text{ bits}}} & \equiv 1 \otimes \underbrace{X \otimes \cdots \otimes X}_{X^{\otimes n}} \ket{\underbrace{\bar{a} + b }_{n+1\text{ bits}}} \ , 
  \label{eq:binarySubPrecise}
\end{align}
meaning the overflow bit supplied to the internal $A$ is not flipped after addition.
When $b>a$, (\ref{eq:binarySub}) yields $a-b$ modulo $2^{n+1}$.
Setting $\ket{c_0}=\ket{p}$ manipulates the subtraction to yield $\ket{\epsilon^{\text{OUT}}-\epsilon^{\text{IN}}-p}$, which is flipped when $\ket{\nu}=\ket{0}$. 
Inversion operators at the end uncompute the constraint, restoring the original configuration.
The net result of the circuit is that $q$ is flipped if and only if $G_s(\mathbf{E},\rho)$ vanishes.

This is all made clearer by considering some simple inputs (taking $q=1$):
\begin{enumerate}[label=(\roman*)]
  \item $\epsilon^{\text{OUT}}=\epsilon^{\text{IN}}=p=\nu =0$:
    The output of $S$ is $\ket{0}^{\otimes (n+1)}$.
    Because $\nu=0$, this gets flipped to $\ket{1}^{\otimes (n+1)}$.
    The $C^n(Z)$ gate is therefore triggered by the physical configuration.
  \item $\epsilon^{\text{OUT}}=\epsilon^{\text{IN}}=0$, $\nu=p=1$:
    The output of $S$ will be $\ket{1}^{\otimes (n+1)}$.
    Because $\nu=1$, this output is not flipped.
    The $C^n(Z)$ gate is therefore triggered by the physical configuration.
  \item $\epsilon^{\text{OUT}}=1$, $\epsilon^{\text{IN}}=p=\nu=0$:
    The output of $S$ will be $\ket{0\cdots01}$, which gets flipped to $\ket{1\cdots 10}$.
    The $C^n(Z)$ gate is consequently not triggered by the unphysical configuration.
\end{enumerate}
More generally, the circuit does not always compute $G_s$ proper, but the $C^n(Z)$ gate is nevertheless triggered exactly when $G_s$ vanishes.

The modification needed for $\mathbb{Z}_{2^n}$ is to omit the overflow bit $h$ and only work with an $n$-bit difference [instead of the $n+1$ bits from (\ref{eq:binarySubPrecise})].
The modification for 1D pure gauge theory is more significant:
For either $g=\text{U(1)}$ or $g=\mathbb{Z}_{2^n}$, physicality is equivalent to saying $E^{\text{OUT}}$ and $E^{\text{IN}}$ are identical.
The oracle is therefore simply constructed using the $\oplus$ gates introduced earlier, bit flips, and a controlled $Z$.

\begin{figure*}
  \begin{center}
    \includegraphics[width=\textwidth]{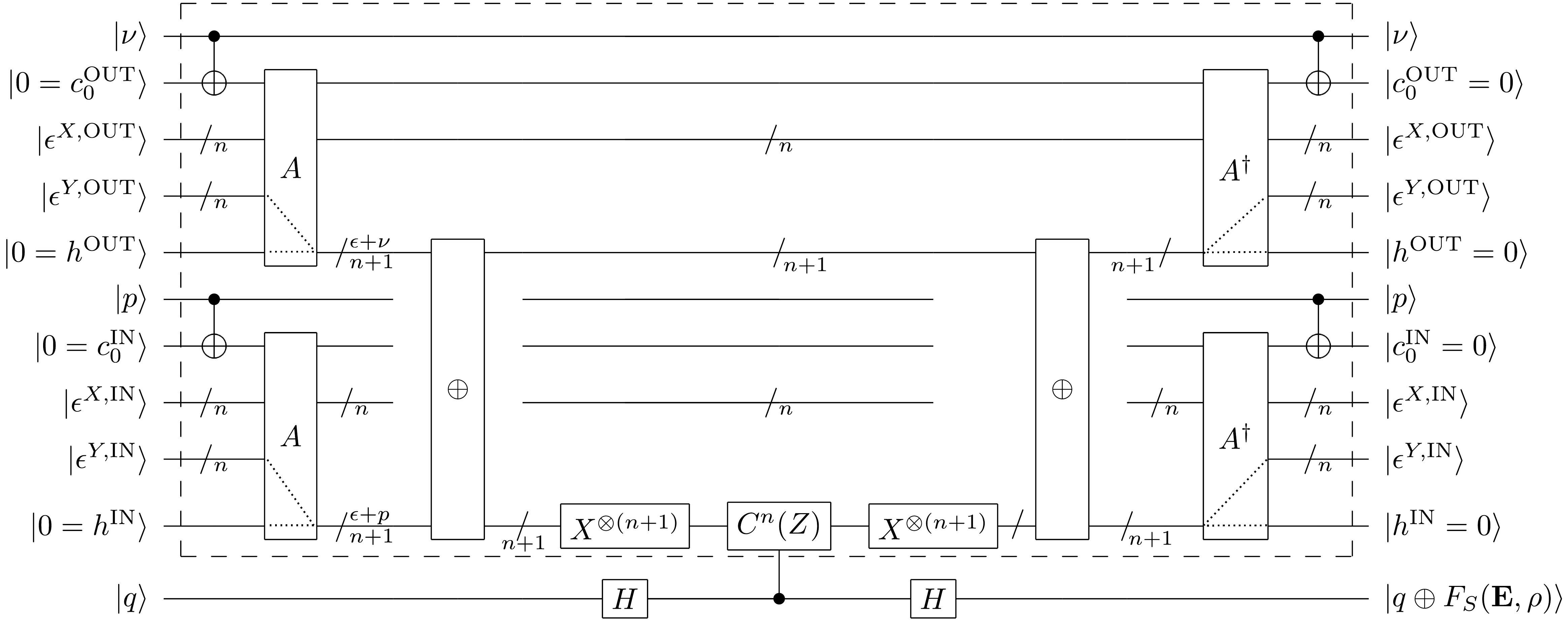}
    \caption{
      A query to the oracle a for U(1) gauge theory in 2D.
      This example accommodates one flavor of Dirac fermion via the occupation numbers $\nu$ and $p$.
      To modify this for $g=\mathbb{Z}_{2^n}$, the overflow bits $h^{\text{OUT,IN}}$ would be removed.
  }
  \label{fig:gaussOracle2D}
\end{center}
\end{figure*}
\subsection{2D U(1): Pure gauge or one Dirac fermion}
In Fig.~\ref{fig:gaussOracle2D}, an oracle for 2D U(1) gauge theory with one Dirace fermion is illustrated.
The inputs to the oracle from the lattice are four $n$-bit electric field states $\ket{\epsilon^{\{x,y\},\{\text{OUT,IN}\}}}$ and occupation numbers $\ket{\nu},\ket{p}$.
Carry bits $\ket{c^{  \text{OUT,IN} }_0}$ and overflow bits $\ket{h^{\text{OUT,IN}}}$ are provided for the adders.
Some lines break over multi-qubit gates, emphasizing that they do not participate in those gates.
Additional work bits could be required by the adder subroutines or by the controlled $Z$.
The result of the circuit is that $q$ is flipped if and only if $G_s(\mathbf{E},\rho)=0$.

In more detail:
The first stage of this circuit involves summing the total out-flux and total in-flux in parallel.
When $\epsilon^{\text{OUT}}_y$ is added to $\epsilon^{\text{OUT}}_x$, a \textsc{cnot} controlled by $\ket{\nu}$ on the incoming carry can have the effect of counting one unit of negative charge in the sense of (\ref{eq:chargedConstraint});
this observation also applies to the in-flux computation with $\ket{p}$ as the control.
The second stage ``adds'' these two results bit-wise.
Third, a phase flip is applied if and only if the bitwise sum was $0\cdots00$;
this is accomplished by flipping the bits, acting with a $C^{n}(Z)$ gate, and then flipping them back.

For this matter scenario, checking (\ref{eq:chargedConstraint}) costs essentially the same as its pure gauge analogue.
The point is that an adder or subtractor for the ``$-\rho$'' term in $G_s$ has been avoided;
more flavors would require more adder subroutines, i.e., many more gates than the four explicit \textsc{cnot}s in Fig. \ref{fig:gaussOracle2D}.
The pure gauge version is obtained by omitting $\ket{\nu}$, $\ket{p}$, and the four \textsc{cnot} gates attached to them.
To modify the 2D oracle for $g=\mathbb{Z}_{2^n}$, the overflow bits $h^{\text{OUT,IN}}$ are omitted from the circuit because they do not need to be calculated.


\subsection{3D U(1): Pure gauge or one Dirac fermion} 

\begin{figure*}
  \begin{center}
    \includegraphics[width=\textwidth]{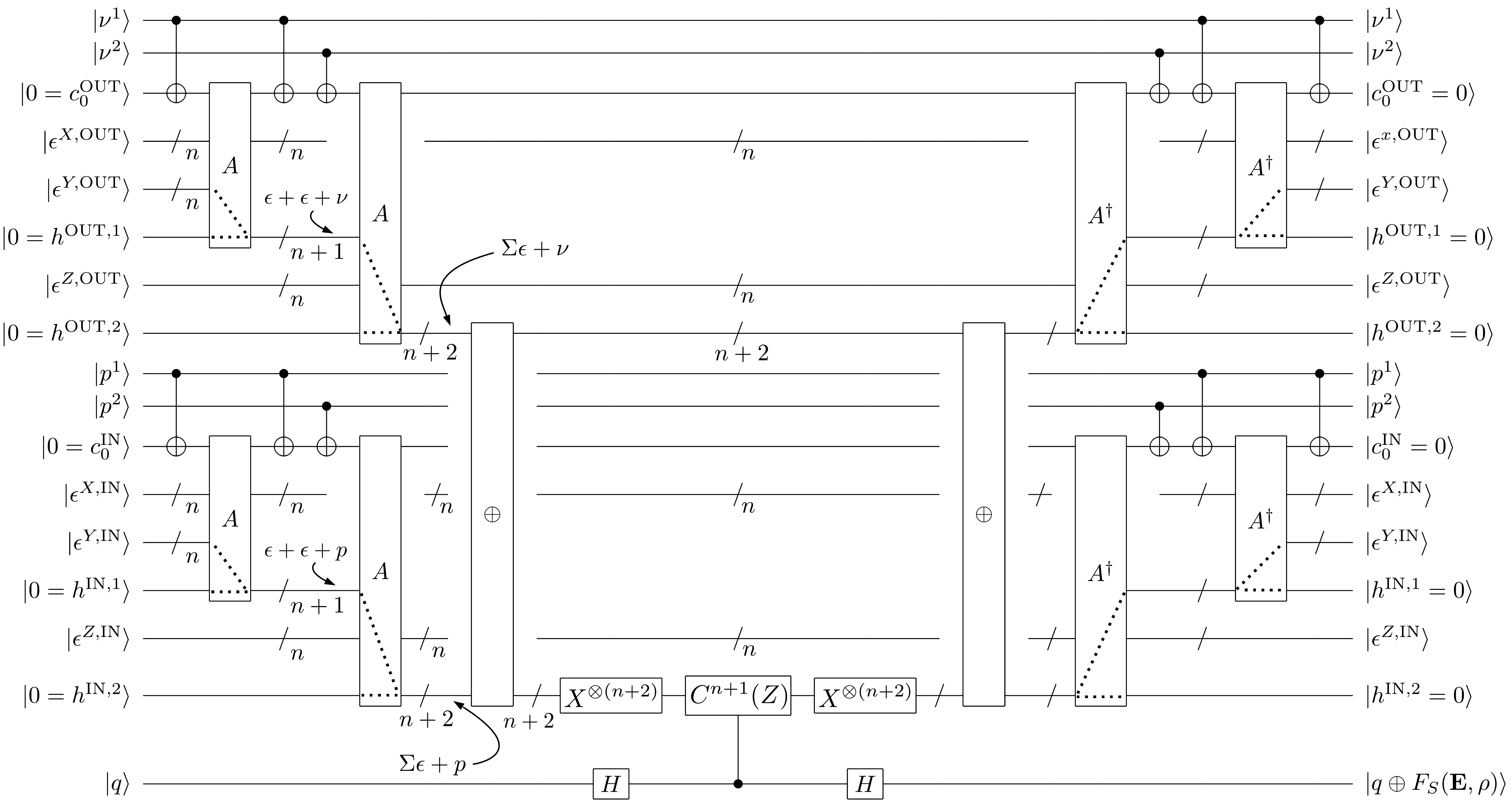}
    \caption{
      A query to the oracle for a U(1) gauge theory in 3D.
      This example accommodates one flavor of Dirac fermion via the occupation numbers $\nu^{1,2}$ and $p^{1,2}$.
      To modify this for $g=\mathbb{Z}_{2^n}$, the four overflow bits $h^{\{\text{OUT,IN}\},\{1,2\}}$ would be removed.
  }
  \label{fig:gaussOracle3D}
\end{center}
\end{figure*}

In Fig.~\ref{fig:gaussOracle3D}, an oracle for 3D U(1) gauge theory is illustrated.
It is very similar to the 2D oracle.
Dirac fermions in 3D have four components, here denoted $\ket{\nu^{1,2}}$ and $\ket{p^{1,2}}$.
Like the previous examples, including one flavor costs marginally more gates than the pure gauge analogue---no more than 12 additional \textsc{cnot}s.
Some of these explicit \textsc{cnot}s serve to reset and reuse the carry bits $c_0^{\text{OUT,IN}}$ a few times throughout the 3D oracle, saving qubits.
Similar comments apply to modifying the algorithm for $g=\mathbb{Z}_{2^n}$ as in 2D:
overflow bits are simply omitted from the computation.

\subsection{\label{subsec:remarks}Remarks on implementation}

While the details of the adder subroutines have thus far been neglected, their implementation generally dominates the oracle gate requirements.
There are several known quantum adder algorithms with varying complexities and resource requirements:
these include ripple-carry
\cite{
  vedral.barenco.eaQuantumNetworks96,
  cuccaro.draper.eaNewQuantum04%
},
quantum Fourier transformation \cite{draperAdditionQuantum00},
carry-lookahead \cite{draper.kutin.eaLogarithmicdepthQuantum05},
carry-save \cite{gossettQuantumCarrySave98},
and later developments of these
\cite{
  thomsen.axelsenParallelOptimization08,
  wang.luo.eaImprovedQuantum16,
  takahashi.tani.eaQuantumAddition10,
  gidneyHalvingCost18%
}.
When simulating something as complicated as a LGT, an $O(1)$ number of ancillary or scratch bits seems reasonable to ask for, in which case the ripple-carry adder \cite{cuccaro.draper.eaNewQuantum04} would be suitable.
The ripple-carry adder (RCA) costs $8n+O(1)$ T gates, and in Table \ref{tab:Tgates} the T counts are given for the examples of the previous section.
\begin{table}
  \caption{\label{tab:Tgates}T counts associated with the arithmetic routines in the oracles for U(1) or $\mathbb{Z}_{2^n}$ gauge theories, coupled to one Dirac fermion, using the RCA method \cite{cuccaro.draper.eaNewQuantum04}.}
\begin{ruledtabular}
\begin{tabular}{c | c}
  Dimension & T count using RCA  (controlled $Z$'s excluded)\\
  \hline
  1D (Fig. \ref{fig:gaussOracle1DCharged})  & $2(8n+O(1))$ \\
  2D (Fig. \ref{fig:gaussOracle2D}) & $4(8n+O(1))$ \\
  3D (Fig. \ref{fig:gaussOracle3D}) & $4(16n+8+O(1))$\\
\end{tabular}
\end{ruledtabular}
\end{table}
Note that a variant of ripple-carry using temporary logical-\textsc{and}s \cite{gidneyHalvingCost18} could be useful if one can further supply $O(n)$ work qubits, because it can halve the T count associated with the oracle's arithmetic.

Finally, the examples only explicitly considered one flavor of Dirac matter.
It is unlikely that several charged species on sites could be accommodated without introducing more adder subroutines to sum $E$'s and $\rho$'s, which would appreciably increase the number of gates in the oracle per the remarks above.
This is already exemplified by the fact that 1D pure gauge oracles require no adder circuits, but one charged species does require an adder.

\section{\label{sec:disc}Discussion}

The previous sections have introduced routines for testing the gauge invariance of wave functions in Abelian LGT simulations by essentially calculating the Gauss law operator.
The simplest interesting examples involving anticommuting matter in $D=1,2,3$, have been worked out explicitly.
It has been shown that the matter content of those theories requires only a small number of gates (independent of $n$) more than the pure-gauge computation, but generally more adders are needed somewhere (increasing the T gate count by $O(n)$).
The remainder of this work revisits the relevance of these routines to a digital quantum LGT simulation, addressing a couple of concerns raised in the Introduction and highlighting the potential applications to error detection. 

The first issue is there is a very real possibility (also pointed out by \cite{stannigel.hauke.eaConstrainedDynamics14,mezzacapo.rico.eaNonAbelianSU15}) that Trotter evolution can generate unphysical components in a state vector due to algorithmic approximation errors.
The first-order Trotter approximation involves replacing time evolution by the full Hamiltonian $H=\sum_j H_j$ with a sequence of small time steps by each $H_j$, and taking the limit of small time steps: 
\begin{equation}
  e^{-i t \sum_j H_j} = \lim_{N_t \rightarrow \infty} \left( \prod_j e^{-i \Delta t \ H_j  } \right)^{N_t}, \quad \Delta t = t/N_t \ .
  \label{<+label+>}
\end{equation}
In general, the individual steps $\exp(-i \Delta t H_j)$ do not commute, so $\prod_j \exp(-i\Delta t\ H_j)\neq \exp(-i\Delta t\sum_j H_j)$, and the state vector suffers $O(\Delta t^2)$ errors dictated by the Baker-Campbell-Hausdorff (BCH) formula.
In a LGT, the usual $H_j$ are all gauge invariant operators, so deviations from the BCH formula ought to preserve Gauss's law.
But to do Trotter evolution on qubits, each $H_j$ would itself need Trotterization down to the level of qubit operations, which is commonly done by decomposing $H_j$ into multi-qubit Pauli operators;
it is well known how to implement Trotter steps once they are in the form $\exp(-i t \ \sigma_{\mu_1}\otimes \sigma_{\mu_2} \otimes \cdots)$.
It is the errors from this second level of decomposition that generally do not commute with Gauss's law.
Hence, at any finite Trotter step size, even a physical initial state and evolution via noiseless gates presents the danger of creating unphysical components in the state. 
Appendix \ref{app:Trotter} expands on these issues for the Schwinger model.
These theoretical errors (as opposed to stochastic) can in principle be quantified, and it may be possible to apply methods from oblivious amplitude amplification
\cite{
  berry.childs.eaSimulatingHamiltonian15,
  berry.childs.eaHamiltonianSimulation15%
}
to help rotate wave functions closer to the physical subspace. 
An algorithm to help ``fix'' a state that has acquired overlap onto the unphysical subspace by rotating it would be extremely useful, and if one exists it will almost certainly require constraint-checking oracles.

The second issue has to do with finite quantum noise---even in the optimistic case of gauge invariant time evolution.
Most of the basis states available to the quantum computer violate Gauss's law, which means there is ample unphysical Hilbert space for a state vector to wander into
\cite{
  stannigel.hauke.eaConstrainedDynamics14,
  kaplan.strykerGaussLaw18%
}.
Error-correcting codes can help to protect against the bit flips that would induce such wandering, but for the foreseeable future it is crucial to save qubits wherever possible.
Therefore routines for verifying gauge invariance would be valuable for filtering out this class of errors.
Such tests could be accept-reject steps either during time evolution or on the final state.

Indeed, a promising application of the oracles is as collective detection mechanisms for bit flip errors in the vicinity of a site;
the gauge invariance condition probes many qubits at once for an error.
To better appreciate this, consider a lattice wave function prepared by acting on a physical initial state with some series of gates.
If bit flip errors are relatively rare throughout the execution, those that do occur are likely to appear as localized Gauss law violations.
This is because a constraint $G_s$ involves quantum numbers from the site $s$ and its $2D$ links, and any single-qubit $X$ (or $Y$) error on them will necessarily change $G_s$.
For multiple $X$ errors on different qubits to look gauge invariant would require they conspire to change the constraint function by compensating amounts.
That is, bit flips can only be overlooked if the error itself corresponds to a gauge invariant operator.
(Note that any function of qubit $Z$s is gauge invariant, so phase errors are invisible to the oracle.)
On small lattices it might be acceptable to simply reject unphysical states, however  the probability of all constraints being preserved in the presence of noise becomes exponentially small with increasing lattice volume.
The benefit is that when all the constraints are found to be satisfied, any unphysical components will have been removed from the wave function.
Alternatively, knowing where or if a lattice wave function has suffered a bit flip error could serve as input for a correction scheme.

Lastly, it should be noted that this paper has focused on the digitization of Abelian LGTs using conventional electric variables, but that is not the only option for quantum simulating LGTs.
In particular, LGTs with finite-dimensional link Hilbert spaces have been introduced
\cite{hornFiniteMatrix81,
  orland.rohrlichLatticeGauge90,
  chandrasekharan.wieseQuantumLink97,
  brower.chandrasekharan.eaQCDQuantum99%
}
and proposed for quantum simulation \cite{
  tagliacozzo.celi.eaOpticalAbelian13,
  tagliacozzo.celi.eaSimulationNonAbelian13,
  pichler.dalmonte.eaRealTimeDynamics16%
}.
Among these are the quantum link models (QLMs).
QLMs have the same local gauge symmetries as the Kogut-Susskind Hamiltonian, but their operator algebra is not identical
\cite{wieseQuantumSimulating14}.
Chandrasekharan and Wiese argue in Ref.~\cite{chandrasekharan.wieseQuantumLink97} that four-dimensional Yang-Mills theory can be obtained from a QLM in the limit of infinite extent of a fifth dimension.
In taking this limit, they relax the Gauss law constraint, which could have some practical advantages;
on the other hand, simulating an extra dimension will present its own challenges.
Another approach to LGT is the prepotential formalism
\cite{
  mathurLoopApproach07,
  raychowdhuryPrepotentialFormulation13,
  raychowdhuryLowEnergy19%
};
non-Abelian generalizations of the oracles in this paper based on prepotentials are also being developed \cite{raychowdhury.strykerTailoringNonAbelian18}.
Many more references for different LGT simulation schemes can be found in the review of Ref.~\cite{dalmonte.montangeroLatticeGauge16}.

\appendix

\section{\label{app:notation}Quantum circuit notation}
This Appendix very briefly introduces quantum circuit notation for the convenience of non-practitioners.

The identification of spin-1/2 states and ``computational basis'' states is $\ket{0} \equiv \ket{\uparrow} $ and $\ket{1}\equiv\ket{\downarrow}$.
The Pauli matrices are often denoted by $X$, $Y$, and $Z$.
Hence, for example, $\hat{Z}\ket{j} = \ket{j} (-)^{j}$ (for $j=0,1$), whereas $\hat{X}\ket{j} = \ket{j\oplus 1} $ (with $\oplus$ denoting addition modulo two).

Single-qubit gates are unitary operations acting on one qubit alone.
For example, the Hadamard gate is
\begin{equation}
  \Qcircuit @C=.3cm @R=.3cm {
& \gate{H} & \qw\\
} \  = \ \frac{1}{\sqrt{2}} \left( \begin{matrix} 1 & 1\\ 1 & -1 \end{matrix} \right) .
\label{<+label+>}
\end{equation}
This paper also refers to T (or ``$\pi / 8$'') gates:
\begin{equation}
  \Qcircuit @C=.3cm @R=.3cm {
& \gate{T} & \qw\\
} \  = \  \left( \begin{matrix} 1 & 0\\ 0 & e^{i \pi /4} \end{matrix} \right) .
\label{<+label+>}
\end{equation}
The initial (final) state corresponds to the left (right) end of the diagram:
\begin{equation}
 \Qcircuit @C=.3cm @R=.3cm {
& \lstick{ \ket{\psi} } & \gate{U_1} & \gate{U_2} & \rstick{\hat{U}_2 \hat{U}_1 \ket{\psi} }\qw\\
} \ .
  \label{<+label+>}
\end{equation}

For two qubits, the ordering of the basis states is usually $\ket{00}, \ket{01}, \ket{10}, \ket{11}$.
To map between mathematical expressions and circuit diagrams, one must choose a convention for how the ordering of the qubits relates to the order of wires.
The first qubit is commonly identified with the top wire, in which case we have, for example,
\begin{center}
  \includegraphics[width=.32\textwidth]{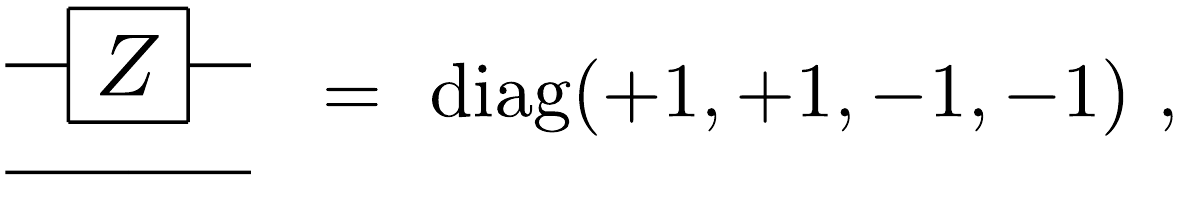} 
\end{center}
and
\begin{center}
  \includegraphics[width=.32\textwidth]{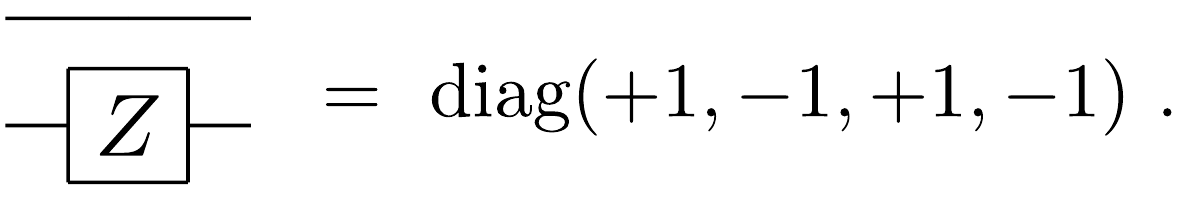} 
\end{center}
The most common two-qubit gate is the ``controlled \textsc{not}'' or \textsc{cnot}, which flips one qubit (the target) when another (the control) is set to $\ket{1}$.
For example,
\begin{center}
  \includegraphics[width=.32\textwidth]{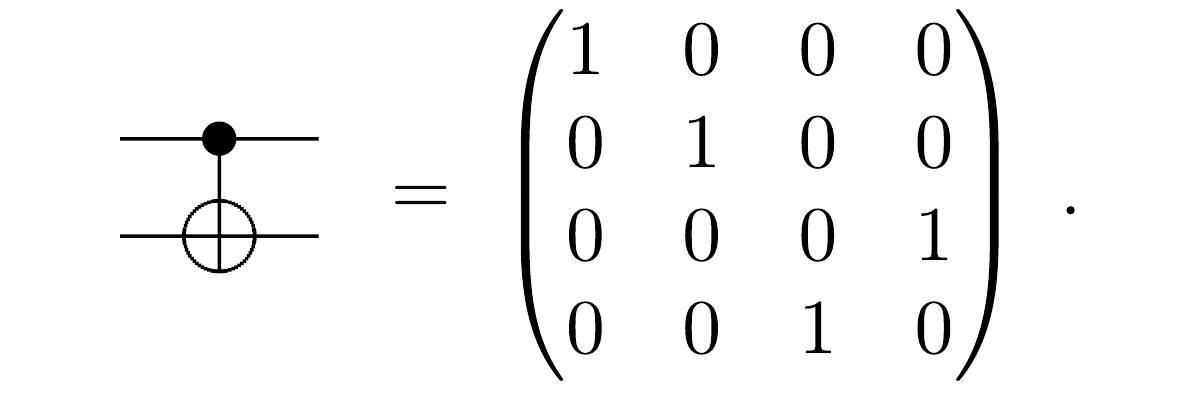} 
\end{center}
A related gate is the ``controlled-$Z$'' or $C(Z)$:
\begin{center}
  \includegraphics[width=.32\textwidth]{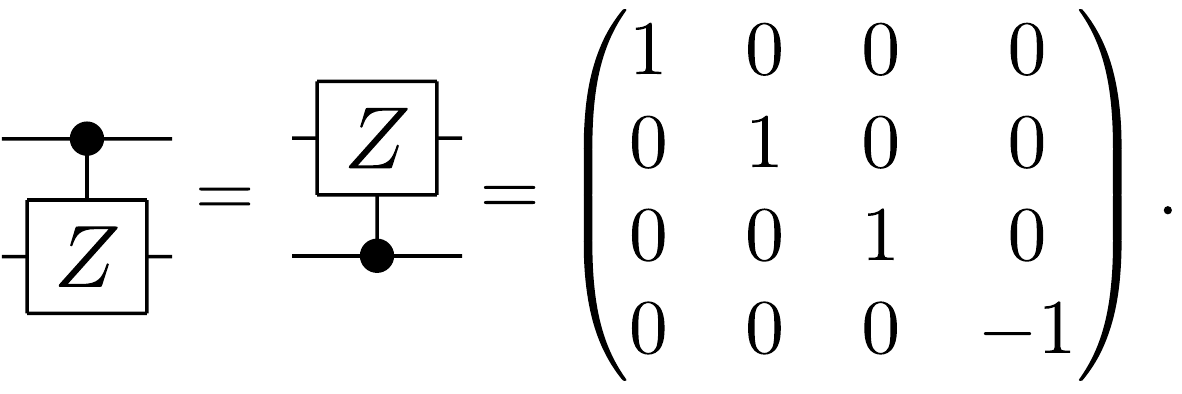} 
\end{center}
The tensor product structure and diagrammatic conventions adopted for two qubits are easily generalized to any number of qubits.
In particular, $C^n(Z)$ generalizes $C(Z)$ by having $n$ control qubits rather than just one.

\section{\label{app:Trotter}Trotter approximation errors in the Schwinger model}
As an example of non-gauge-invariant errors, consider a Trotter step in the Schwinger model [a U(1) gauge theory in 1D]
\cite{schwingerGaugeInvariance62}.
Discretizing space with staggered fermions
\cite{kogut.susskindHamiltonianFormulation75},
employing periodic boundary conditions,
and using a Jordan-Wigner transformation
\cite{jordan.wignerUeberPaulische28}, a rescaled Hamiltonian for this theory is
\cite{
  martinez.muschik.eaRealtimeDynamics16,
  klco.dumitrescu.eaQuantumclassicalComputation18
}
\begin{align}
  \begin{split}
    \hat{H} &=  x \sum_{s=0}^{2 N_{\text{ph}}-1}[\sigma^{-}(s) \hat{U}(s) \sigma^{+}(s+1) + \text{H.c.} ]\\
    &\quad + \sum_{s=0}^{2 N_{\text{ph}}-1}[\hat{E}(s)^2 + \frac{\mu}{2}(-)^s \hat{Z}(s)] \ .
  \end{split}
  \label{eq:hamSchwinger}
\end{align}
Here, $N_{\text{ph}}$ is the number of physical sites, $x,\mu$ are parameters, $E(s),U(s)$ are the operators introduced in section \ref{sec:bg}, and $\sigma_n^{\pm}$ change fermionic occupation numbers.
Trotter errors induced by decomposing the electric energy or the mass term into separate steps by individual Pauli operators (if there are any such errors) will not affect Gauss's law.
The troublesome part of the Hamiltonian is the hopping term,
\begin{align}
  \hat{H}_{\text{h}} =& \sum_{s=0}^{2 N_{\text{ph}}-1} \hat{H}_\text{h}(s)\ ,\\
  \hat{H}_\text{h}(s)=&\left[\sigma^{-}(s) \hat{U}(s) \sigma^{+}(s+1) + \sigma^{+}(s) \hat{U}^{\dagger}(s) \sigma^{-}(s+1) \right] .
  \label{eq:hamHop}
\end{align}
If the links are given a cutoff $n=1$, then $\hat{H}_\text{h}(s)$ decomposes into four Pauli operators,
\begin{align}
  \hat{H}_\text{h}(s)|_{n=1} = \frac{1}{4}(XXX+XYY-YYX+YXY) \ .
  \label{eq:hamHopPauli1}
\end{align}
The notation here uses implicit tensor products where the first (last) operator acts on site $s$ ($s+1$), and operators in the middle act on the link.
It turns out that all the operators in (\ref{eq:hamHopPauli1}) mutually commute, so for this special case elementary methods can be used to implement $\exp(-i \Delta t\ x H_\text{h}(s))$. 
When the cutoff is $n=2$, however, $H_\text{h}(s)$ is the sum of 12 Pauli operators,
\begin{align}
  \begin{split}
    \hat{H}_\text{h}(s)&|_{n=2} = \\
    &\frac{1}{4}[XIXX+XIYY-YIYX+YIXY] \\
    &+ \frac{1}{8}\left[XXXX+XYYX-XXYY+XYXY\right.\\
    & \qquad \left. + YXYX-YYXX+YXXY+YYYY\right] \ . 
  \end{split}
  \label{eq:hamHopPauli2}
\end{align}
These operators do not all commute with each other, and the Trotter step obtained by compounding 12 elementary rotations generally differs from $\exp(-i\Delta t \ x H_\text{h}(s))$ by errors that break Gauss's law.
The situation worsens with increasing $n$, so na\"{i}ve Trotterization is destined to create unphysical components in a state vector.
Alternatives to Trotterization, such as quantum walks
\cite{
  berry.childsBlackboxHamiltonian12,
  childs.wiebeHamiltonianSimulation12%
},
could also face the same problems with gauge invariance, but whether the situation is better or worse was not investigated (see also \cite{marquez-martin.arnault.eaElectromagneticLattice18}).

\begin{acknowledgments}
  I thank David B.~Kaplan, Natalie Klco, Indrakshi Raychowdhury, Alessandro Roggero, and Martin J.~Savage for helpful discussions and feedback on the manuscript.
  I also thank Natalie Klco and Alessandro Roggero for directing me to important references.
  This material is based upon work supported by the National Science Foundation Graduate Research Fellowship Program under Grant No.~DGE-1256082.
  Any opinions, findings, and conclusions or recommendations expressed in this material are those of the author(s) and do not necessarily reflect the views of the National Science Foundation.
  This work was also supported in part by DOE Grant No.~DE-FG02-00ER41132.
\end{acknowledgments}

\bibliography{abelianGauss.v21.bib}

\end{document}